\documentclass[12pt]{article}
\usepackage{amsmath}
\usepackage{graphicx}
\usepackage[table]{xcolor}
\usepackage{anyfontsize}
\usepackage[authoryear]{natbib}
\usepackage[tablewithin=none,figurewithin=none,singlelinecheck=false]{caption}
\usepackage[referable]{threeparttablex}
\usepackage[utf8]{inputenc}
\usepackage{longtable}
\usepackage{moresize}
\usepackage{float}
\usepackage{natbib}
\usepackage{hyperref}
\usepackage{booktabs}
\usepackage{tabularx}
\usepackage[hmargin=3cm]{geometry}
\usepackage{tikz}
\usetikzlibrary{matrix,calc}
\usepackage{array}
\usepackage{multirow}
\usepackage{float}
\usepackage{authblk}
\usepackage{geometry}
\usepackage{longtable,booktabs}
\usepackage{lscape}
\newcommand\MyBox[2]{
  \fbox{\lower0.75cm
    \vbox to 1.7cm{\vfil
      \hbox to 1.7cm{\hfil\parbox{1.4cm}{#1\\#2}\hfil}
      \vfil}%
  }%
}

\title{Feasibility of Mental Health Triage Call Priority Prediction Using Machine Learning}

\author{Rajib Rana$^1$, Niall Higgins$^{1,2}$, Kazi Nazmul Haque$^1$, John Reilly$^3$, Kylie Burke$^{4,5,6}$, Kathryn Turner$^4$, Terry Stedman$^2$ }
\date{%
    $^1$University of Southern Queensland, Australia\\%
    $^2$Mental Health and Specialist Services, West Moreton Health, Brisbane, QLD 4076, Australia\\%
     $^3$Mental Health Alcohol and Other Drugs Branch, Clinical Excellence Division, Queensland Health, Brisbane, QLD 4029, Australia\\%
       $^4$Metro North Mental Health, Metro North Health, Brisbane, QLD 4029, Australia\\%
       $^5$School of Psychology, The University of Queensland, Brisbane, QLD 4072, Australia\\%
       $^6$Australian Research Council’s Centre of Excellence for Children and Families over the Life Course, Brisbane, QLD 4068, Australia\\%
}

\begin{document}

\maketitle

\begin{abstract}
Ensuring accurate call prioritisation is essential for optimising the efficiency and responsiveness of mental health helplines. Currently, call operators rely entirely on the caller's statements to determine the priority of the calls. It has been shown that entirely subjective assessment can lead to errors. Furthermore, it is a missed opportunity not to utilise the voice properties readily available during the call to aid in the evaluation. Incorrect prioritisation can result in delayed assistance for high-risk individuals, resource misallocation, increased mental health deterioration, loss of trust, and potential legal consequences. It is vital to address these risks to guarantee the reliability and effectiveness of mental health services. This study delves into the potential of using machine learning, a branch of Artificial Intelligence, to estimate call priority from the callers' voices for users of mental health phone helplines. After analysing 459 call records from a mental health helpline, we achieved a balanced accuracy of 92\%, showing promise in aiding the call operators' efficiency in call handling processes and improving customer satisfaction.
\end{abstract}

\section{Introduction}
Mental illness is a leading cause of global burden of disease accounting for 13\% disability adjusted life-years \citep{Vigo2022}. In recent decades the scale of mental illness has risen even prior to the COVID-19 pandemic \citep{Botha2023}. This presents additional challenges due to the current shortage of mental health clinicians and psychiatrists \citep{Nguyen2023}. Due to the significant morbidity and mortality associated with mental illness, there is an urgent need to identify callers to helplines who have a high level of distress and consequently have an increased need to be seen by a clinician who can offer interventions for treatment.

Mental illness is multifactorial and to understand it requires not only measurable biometric data but also consideration of the complex social and historical context that impacts a person’s health outcomes \citep{Villongco2020}. Many psychiatric illnesses incorporate a voice component into their diagnostic criteria. For instance, disorganised speech is a key feature in diagnosing schizophrenia, while symptoms such as excessive talkativeness or pressure to keep talking characterise manic episodes in diseases like Bipolar I and II \citep{Perlini2012}. Voice computing and Machine Learning (ML) hold significant promise for improving identification of those with high priority mental health needs by supporting the clinician with their assessment of a caller’s need for care \citep{Low2020}.

Efficiently managing incoming calls is crucial for maintaining high levels of patient safety and operational efficiency in modern call centres. This was especially evident during the recent pandemic where there was a rapid increase in the number of these centres established in multiple countries to deliver psychological services, Covid-19-related information, consultation, and triage services \citep{Jahromi2024}. The majority of call centre operators were volunteers with a variety of skills including training in mental health. Among other factors, lessons learned indicated that sufficient human resources, adequate financial and physical resources, and platforms for objective clinical assessment of the callers should be considered for successful management of a call centre. Currently, in Australia, confidential mental health telephone triage call centres handle a diverse array of calls, from routine inquiries to urgent issues requiring immediate attention. These call centres provide care for a range of needs that include mental health, domestic violence, sexual assault and LGBTQIA+ support services.  Operating 24 hours a day, they are critical for the early detection of mental illness and to prioritise access to early treatment and reduce the unnecessary use of resources \citep{Sands2015, Thomas2021}. Whilst triage has its origins in emergency medicine, there are relatively few studies that have examined the use of triage in access to mental health care \citep{Watson2022}. 

Mental health triage decision support aids assist clinicians with responding appropriately to the caller's need and there is some evidence to indicate that they improve clinical judgement \citep{Downey2014, Tanner2014}. Contemporary clinical decision support systems for call centre triage are mainly used to confirm clinical assessments and perceived as a support rather than the previously held views of limiting professional autonomy \citep{Holmstrom2020}. These systems are typically either knowledge-based medical questions, where decisions from caller answers are made according to guidelines, protocols or clinical expertise, or they are data driven using statistical or machine learning models \citep{Michel2024}. Currently, there is a pressing need to improve clinical capability of health professionals in mental health call centres as well as for those who are not mental health clinicians \citep{Briggs2021}. 

The most obvious obstacle in mental health telephone triage is the ability to use visual clues which are an integral part of the decision making that will determine the care that will be afforded to the caller. Imminent suicidal risk is the most important triage issue. The clinician is compelled to listen for signs of anxiety or depression for example even if the caller does not specifically state these symptoms. They are limited by their ability to communicate with the caller effectively enough for the clinician to make an assessment of priority for potentially life-threatening behaviour by a caller. The introduction of AI is well-suited to support triage for identification of these sporadic occasions as most calls are less critical and information provided by the caller is gathered for referral. The substantial impact on health caused by mental illness and distress has driven the research conducted here towards more advanced computational approaches for prediction of priority need level and personalised treatment. This paper investigates the feasibility of implementing a machine learning or deep learning-based call priority prediction system that predominantly uses vocal properties of the caller's voice and not what is actually said. It is designed to assist the clinician to identify the priority of mental health need for a caller and support the triage decision.

\section{Related Work}
The application of machine learning in call centre operations has gained significant attention in recent years. These techniques offer a data-driven approach to address the limitations of traditional call prioritisation methods. It is known that linguistic markers have predictive value in psychiatric assessments, and language has the potential to be a biosocial marker for conditions such as psychosis \citep{Palaniyappan2021}. Voice computing, when applied to callers to a helpline, employs statistical learning. It uses voice features as input to explain specific outputs related to symptoms of mental illness. The method requires a learning period and relies on training datasets for success. We discuss the existing literature by grouping it into two parts: the first part discusses the existing literature where machine learning techniques have been used for call priority. The second part discusses commercially available products/applications for assisting with call priority. 

\subsection{Machine Learning in Call centres}
% Several studies have explored using machine learning to improve call centre operations. \citep{gans2003telephone} provided a comprehensive review of call centre management and highlighted the potential of data-driven approaches. 

Most of the existing studies use call transcriptions to determine the call priorities. For example, Ceklic et al. \cite{ceklic2022ambulance} show that natural language processing on emergency medical text has a high predictive ability in identifying vehicle crashes that require the fastest ambulance response and those that do not. A retrospective cohort study used data from 2014 to 2016 traffic crashes attended by emergency ambulances in Perth, Western Australia. Machine learning algorithms were used to predict the need for a lights and sirens (L\&S) response or not. The features were the Medical Priority Dispatch System (MPDS) determinant codes and EMD (Emergency Medical Department) text. EMD text was converted for computation using natural language processing (Bag of Words approach). Machine learning algorithms were used to predict the need for an L\&S response, defined as where one or more patients (a) died before hospital admission, (b) received lights and sirens L\&S transport to the hospital, or (c) had one or more high-acuity indicators (based on an a priori list of medications, interventions or observations. There were 11,971 traffic crashes attended by ambulances during the study period, of which 22.3 \% were retrospectively determined to have required an L\&S response. The model with the highest accuracy was using an Ensemble machine learning algorithm with a score of 0.980 (95 \% CI 0.976–0.984). 

Similarly, Anthony et al. \cite{anthony2021feasibility} presents the application of machine learning for classifying time-critical conditions, namely sepsis, myocardial infarction and cardiac arrest, based on transcriptions of emergency calls from emergency services dispatch centres in South Africa. The data set comprised manually transcribed emergency call conversations from various EMS contact centres in South Africa. The data set consisted of an original data set of 93 examples, and the authors present results from the application of four multi-class classification algorithms: Support Vector Machine (SVM), Logistic Regression, Random Forest and K-Nearest Neighbor (kNN). Authors achieved 95\% accuracy when predicted on unseen data.

Again, Defilippo et al.~\cite{defilippo2024leveraging} define and implement an AI-based module to manage patients' emergency code assignments in emergency departments. It uses historical data from the emergency department to train the medical decision-making process. Data containing relevant patient information, such as vital signs, symptoms, and medical history, accurately classify patients into triage categories. Their methodology leverages graph representation learning, which is applied to networks through graph neural networks. Graph Representation Learning (GRL) is a method for encoding the structural information of a graph into low-dimensional vectors. Their system receives patient data as input to build the patient similarity network, where each node is a patient, and the weighted edges model the similarity among them. Their system implementation uses four known measures for evaluating similarity: cosine similarity, Euclidean, Manhattan, and Minkowsky distances. Experimental results demonstrate that the proposed algorithm achieved high accuracy, outperforming traditional triage methods.

Kanaan et al.~\cite{kanaan2023methodology}, however, used both audio and text. They demonstrate through their results that it is possible, with the right selection of algorithms, to predict if the call will result in a serious injury with a 71\% accuracy based on the caller's speech only. The data used for this work is one week's worth of emergency call recordings provided by the French SDIS 25 firemen station located in the Doubs.

We only found a small number of studies using call records to estimate call priority. Here, we include a few other studies which are loosely connected with the proposed study. For example, Rosa~\cite{perez2023emotion} et al. use a methodology capable of identifying the emotional state of the patient is and able to classify it in real time. The authors used a range of software development tools to collect data from the coordinates of identified facial features. Finally, a model was applied using the SVM vector machine for the classification and detection of evidence extracted from emotions. As a result, the mood of the patients is evaluated, and the necessary measures are taken to improve online care.

Another study, conducted by Arngeir~\cite{berge2023designing} addresses how we can design for
control in human-AI collaboration in order to enhance rather than
replace human decision-making processes.

\subsection{Commercially Available Products}
A few commercially available products also employ machine learning and AI for call priority prediction. {\bf Freshdesk's} \cite{Freshdesk} sentiment analysis feature leverages AI to assess the emotional tone of customer interactions. A support ticket is allocated to the call and stored in the management system. Tickets are then tagged with sentiments like positive, neutral, or negative, enabling agents to prioritise tickets based on the urgency indicated by the sentiment of the caller. For example, a ticket with a negative sentiment might be prioritised over a neutral or positive one, even if the initial categorisation was of lower priority, to address the customer's emotional state effectively. 

{\bf Genesys Cloud} \cite{GenesisCloud} allows the call management system to set the priority of interactions using the "Set Priority" action within their Architect in-queue flows. This enables dynamic prioritisation based on predefined criteria, ensuring that high-priority calls are addressed promptly while they wait in the queue.

Although not calls, another commercially available product {\bf Zendesk} \cite{Zendesk} can prioritise customer interactions using a combination of automation, AI, and custom configurations to enhance the efficiency and responsiveness of call centres. Automated ticket prioritisation is achieved through custom triggers and automation that sets priority levels based on specific conditions, such as keywords or customer segments. The use of AI and machine learning further refine this process by predicting ticket priority using historical data. Custom fields and views allow agents to organise tickets based on priority, ensuring high-priority issues are addressed promptly.

\subsection{Summary of Related Work}
We present a summary of the key studies in Table~\ref{tab:relatedwork}.
\begin{enumerate}
    \item We find only a few studies that consider improving call prioritisation through the use of machine learning.
    \item Most studies use audio transcription, with only studies using audio and text separately. This leaves tremendous opportunities to explore audio features to understand call priority, given speech is a crucial affective display.
    \item Most studies use shallow machine learning methods, leaving room for advanced deep learning techniques.
    \item Only one existing product, Freshdesk, uses sentiment analysis on audio transcription, leaving room for capitalising on affective information that can be extracted through speech - a crucial conveyer of emotion and attitude. 
\end{enumerate}
% Table \ref{tab:relatedwork} summaries relevant studies that have used machine learning or deep learning to determine the priority or risk of callers in call centres, as well as commercially available products for this purpose.

%\begin{table}[hp]
%   \centering
%    \caption{Summary of studies and products on call priority prediction using machine learning.}
%    \label{tab:relatedwork}
%   % \begin{tblr}{@{}lllX[c,valign=b]]X[c,valign=b]]@{}}
%    \begin{tabularx}{1\textwidth} { 
%  | >{\raggedright\arraybackslash}X 
%  | >{\centering\arraybackslash}X 
%  | >{\raggedleft\arraybackslash}X 
%   | >{\raggedleft\arraybackslash}X  
%  | >{\raggedleft\arraybackslash}X  
%  }\hline
        %\toprule
\begin{ThreePartTable}
\begin{longtable}{|p{2.5cm}|p{3cm}|p{1.8cm}|p{4.7cm}|p{3cm}|}
%\captionsetup{margin=-8pt}
\caption{Summary of studies and products on call priority prediction using machine learning.} 
\label{tab:relatedwork} \\ 
\hline
\textbf{Study/\newline Product} & \textbf{Method} & \textbf{Accuracy} & \textbf{Dataset Used} & \textbf{Predicted \newline Observation}\\ \hline
\endfirsthead

\multicolumn{5}{l}
{Table \thetable\ continued from previous page} \\
\hline
\textbf{Study/\newline Product} & \textbf{Method} & \textbf{Accuracy} & \textbf{Dataset Used} & \textbf{Predicted \newline Observation} \\ \hline
\endhead

\hline \multicolumn{5}{r}{{Continued on next page}} \\
\endfoot

\hline
\endlastfoot       
%        \textbf{Study/Product} & \textbf{Method} & \textbf{Accuracy} & \textbf{Dataset Used} \\ \hline
       \textbf{Ceklic 2022} \cite{ceklic2022ambulance} Retrospective cohort study
        & \small{Natural language processing on emergency medical text. Ensemble, K-nearest \newline neighbours \newline (k-NN), Naïve Bayes, Neural Network and Support Vector Machine} 
        & 98\% 
        & \small{Text sent to paramedics en-route to the traffic crash scene by the emergency medical dispatcher (EMD), in combination with dispatch codes for all traffic crashes attended by SJ-WA paramedics over the study period in the Perth metropolitan area.} 
        & \small{Which incidents require a lights and sirens \newline ambulance \newline response}\\ 
        \hline
        
        \textbf{Anthony 2021} \cite{anthony2021feasibility} \newline Retrospective cohort study 
        & \small{Support Vector Machine (SVM), Logistic Regression, Random Forest and K-Nearest Neighbor (kNN).} 
        & 95\% 
        & \small{The data set comprised of manually transcribed 93 emergency call conversations from various EMS contact centres in South Africa. The transcriptions were originally collected in various South African languages, namely: English, Afrikaans, Zulu and Sesotho.} 
        & \small{Accurately identify critical conditions such as sepsis and cardiac arrest, based upon \newline transcriptions of emergency calls.} \\ \hline
        %\citep{gans2003telephone} & Logistic Regression, Decision Trees & 78\% & Simulated Call Center Data \\\hline
        % \citep{fleury2015svm} & Support Vector Machine (SVM) & 83\% & Health Smart Home Call Data \\ \hline
       % \citep{hussain2018hybrid} & Hybrid ANN and Fuzzy Logic & 88\% & Telecommunications Service Data \\\hline
        %\citep{kim2017customer} & Deep Belief Network (DBN) & 91\% & E-commerce Customer Support Data \\ \hline
        %\citep{wang2018speech} & Speech Emotion Recognition with CNN & 87\% & Call Center Audio Data \\ \hline
        %\citep{gupta2019sentiment} & Sentiment Analysis with LSTM & 89\% & Social Media Interactions and Call Logs \\ \hline
        \textbf{Defilippo 2024} \cite{defilippo2024leveraging} \newline Retrospective cohort study
        & \small{Graph Neural Network} 
        & \small{"a high accuracy in predicting triage code."} 
        &  \small{Publicly available historical data from the Kaggle platform containing relevant patient information, such as vital signs, symptoms, and medical history.} 
        & \small{Accurately identifying a triage code based on clinical presentation of a patient to an emergency department.}\\ 
        \hline
        
        \textbf{Kanaan 2023} \cite{kanaan2023methodology} \newline Retrospective cohort study
        & XGBoost, LSTM, TextCNN, XLM-RoBERTa, CamemBERT 
        & 71\% 
        &  One week of emergency call recordings provided by the French SDIS 25 firemen station 
        & To predict if the call will result in a serious injury based on the caller’s speech only and assigning an appropriate priority to their call. \\ \hline
        % \citep{besse2020machine} & Machine Learning Ensemble & 88\% & Insurance Call Data \\
        % \citep{chen2019application} & LSTM with Attention Mechanism & 90\% & Healthcare Call Data \\
        %\citep{liu2019predicting} & Gradient Boosting Decision Trees & 87\% & Retail Customer Service Data \\ \hline
        %\citep{li2020deep} & Deep Reinforcement Learning & 92\% & Telecom Customer Support Data \\ \hline
        
        \textbf{Freshdesk} \cite{Freshdesk}
        & \small{AI-driven Priority Assignment} 
        & \small\textbf{N/A} 
        & \small\textbf{Proprietary Customer Data} 
        & \small{To assess the emotional tone of customer interactions.} \\ 
        \hline
        
        \textbf{Genesys Cloud} \cite{GenesisCloud}
        & \small{Predictive Routing with AI} 
        & \small\textbf{N/A} 
        & \small\textbf{Proprietary Customer Data} 
        & \small{To set priority of interactions while callers waiting in a queue.} \\ 
        \hline
        
        \textbf{Zendesk} \cite{Zendesk}
        & \small{Machine Learning-based Prioritization} 
        & \small\textbf{N/A} 
        & \small\textbf{Proprietary Customer Data} 
        & \small{To ensure high priority issues are addressed promptly.} \\ \hline
        %\bottomrule
      
%    \end{tabularx}
\end{longtable}
\end{ThreePartTable}

\section{Experimental Setup and Results}
\subsection{Dataset} 
We gathered call data from a mental health access line into a large public mental health hospital providing services to people with serious mental illness in Australia. The help line is a confidential mental health telephone triage service that acts as the first point of contact to public mental health services. The study was approved under the Public Health \textit{Act} with ethical approval (project ID: 61948) and all calls were manually de-identified before audio analysis. The service operates 24/7, and during this study calls were initially answered by a trained administrative officer who collected some initial personal information to determine the best course of action. If the situation is urgent, callers will be connected to a clinician for immediate assistance. If not urgent, an appointment will be scheduled for a clinician to call the individual back at a convenient time. All mental health clinicians are highly trained and experienced professionals, dedicated to providing the best care possible. They give callers the opportunity to discuss their concerns and work with them to develop a management plan, which may include further mental health assessment and treatment. This evaluation process, also called triage, typically takes between 15 and 30 minutes. To help guide the management plan, the clinician determines a triage level based on the triage scale shown in Table~\ref {tab:triage}, that was developed in 2015 by Sands et al. from the United Kingdom\citep{Sands2015}. They key idea of this paper is to use historical triage records to train a machine learning model to able to predict the triage score for a new call.

\begin{ThreePartTable}
\begin{longtable}{|p{2.5cm}|p{2.8cm}|p{2.8cm}|p{3.3cm}|p{3cm}|}
%\captionsetup{margin=-8pt}
\caption{UK Mental Health Triage Scale \cite{Sands2015}} \label{tab:triage} \\ \hline
\small \textbf{Triage Code\newline /description} 
& \small\textbf{Response Type/ Time \newline to Contact} 
& \small\textbf{Typical \newline Presentations} 
& \small\textbf{Mental Health Service Action/\newline Response} 
&\small \textbf{Additional \newline Actions to be Considered}  
\\ \hline

\endfirsthead

\multicolumn{5}{l}%
{ Table \thetable\ continued from previous page} \\
\hline
\small\textbf{Triage Code\newline /description} 
& \small\textbf{Response Type/ Time \newline to Contact} 
& \small\textbf{Typical \newline Presentations} 
& \small\textbf{Mental Health Service Action/\newline Response} 
&\small \textbf{Additional \newline Actions to be Considered}  
\\ \hline
\endhead

\hline \multicolumn{5}{r}{{Continued on next page}} \\
\endfoot

\hline
\endlastfoot

\cellcolor[HTML]{CC0000} \centering \small \textbf{A} \newline \textbf{Emergency} 
& \small \raggedright \color{red}{\textbf{IMMEDIATE} \newline \textbf{EMERGENCY} \newline \textbf{RESPONSE}}  
&  \small  {Current actions endangering self or others \newline Overdose / suicide attempt \newline Violent behavior \newline Possession of a weapon } 
& \small {Triage/clinician to notify ambulance, police and/or mental health services \newline Telephone support }
& \small {Keeping caller on the phone until emergency services arrive \newline Ensure own safety}
\\ \hline

\cellcolor[HTML]{FF9900}\centering \small \textbf{B} \newline \textbf{Very high risk to self or others}  
& \small \raggedright {\color{red}{\textbf{WITHIN 4 HOURS}}} \newline \textbf{Urgent mental health response}  
& \small{Acute suicidal ideation or risk of harm to others \newline High risk of self-harm or aggression \newline Psychotic behavior associated with risk \newline Highly distressed person (e.g., severe agitation or panic) \newline No other suitable agency available \newline Significant worsening of behavior/mental state} 
& \small{Crisis Team/ \newline Liaison/ \newline Triage clinician to notify psychiatrist \newline Arrange face-to-face assessment} 
& \small{Refer additional support and services \newline Triage clinician to notify appropriate services \newline Contact point of situation changes}
\\ \hline

\cellcolor[HTML]{FFCC00}\centering \small \textbf{C}  \newline \textbf{High risk of harm to self or others} \newline \textbf{Significant distress} \newline \textbf{Needs urgent supports}  
& \small \raggedright {\color{red}{\textbf{WITHIN 24 HOURS}}} \newline \textbf{Urgent mental health response}  
& \small{Suicidal ideation with plan or ongoing suicidal thoughts \newline Person unable to care for self or neglecting essential needs \newline Psychotic behavior, distress, agitation \newline Severe anxiety or panic \newline High distress in vulnerable person \newline Risk of abuse or harm in care home} 
& \small{Crisis Team/Liaison/ \newline Community Mental Health Team (CMHT) face-to-face assessment }
& \small{Contact same day with a review of appropriate actions \newline Liaison with psychiatrist \newline Follow up assessment of situation changes} \\ 
\hline

\cellcolor[HTML]{66CC33} \centering \small \textbf{D}  \newline \textbf{Moderate risk of harm to self or others} \newline \textbf{Needs mental health response}  
& \small \raggedright {\color{red}{\textbf{WITHIN 72 HOURS}}} \newline \textbf{Urgent mental health response}  
& \small{Significant patient/case distress associated with mental health condition \newline Some risk of self-harm or aggression \newline Disorientation, confusion \newline Possible risk of deterioration without intervention }
& \small{Liaison/CMHT face-to-face assessment }
& \small{Send appropriate information to referrer \newline Situation review within 72 hours} \\ 
\hline

\cellcolor[HTML]{66CCFF}\centering  \small \textbf{E}  \newline \textbf{Low risk of harm to self or others} \newline \textbf{Requires mental health support within days}  
& \small \raggedright {\color{red}{\textbf{WITHIN 4 DAYS}}} \newline \textbf{Non-urgent mental health response}  
& \small{Requires specialist mental health intervention \newline Current diagnosis indicating further assessment \newline Risk of deterioration without intervention \newline Low level suicidal ideation \newline Anxiety, depression, other mental health symptoms impacting daily life }
& \small{Out-patient clinic or CMHT face-to-face assessment }
& \small{Telephone support and advice \newline Triage clinician to review if situation changes} \\ 
\hline

\centering \small \textbf{F} \newline \textbf{Referral to GP or other health/social care provider} 
& \small \textbf{Referral to appropriate service within 7 days} 
& \small{Other services (outside mental health) more appropriate to current situation or need} 
& \small{Triage clinician to advise on appropriate service} 
& \small{Assist and/or facilitate transfer to appropriate service \newline Telephone support and advice} \\ \hline

\centering \small \textbf{G} \newline \textbf{Advice, consultation, information}  
& \small \textbf{Advice or information given, no further mental health intervention required}
& \small{Patient or carer requiring advice or information only \newline Self-managing or awaiting further clinical assessment }
& \small{Triage clinician to provide advice and information}
& \small{Consider courtesy follow-up call if appropriate \newline Telephone support and advice} \\ 
\hline
\end{longtable}
\end{ThreePartTable}
\begin{table}[H]
    \centering
    \begin{minipage}{0.3\textwidth}
        \centering
        \begin{tabular}{|c|c|}
            \hline
            \textbf{Categories} & \textbf{Triage Level} \\ \hline
            High Priority & A, B, C, D \\ \hline
            Low Priority & E, F, G \\ \hline
        \end{tabular}
        \caption{Triage Levels}
        \label{tab:triage_levels}
    \end{minipage}
     \hfill
    \begin{minipage}{0.55\textwidth}
        \centering
        \begin{tabular}{|c|c|c|}
            \hline
            \textbf{} & \small \textbf{Train Samples} & \small \textbf{Test Samples} \\ \hline
            High Priority & 171 & 38 \\ \hline
            Low Priority & 206 & 44 \\ \hline
            Total & 377 & 82 \\ \hline
        \end{tabular}
        \caption{Distribution of Train and Test Samples}
        \label{tab:train_test_samples}
    \end{minipage}%
       
\end{table}
\subsection{Pre-processing}
We collected a total of $459$ call records from the 1300 MH CALL service. As shown in Table~\ref {tab:triage}, triage scores span a scale marked A to G with A assigned to indicate highest call priority and G is assigned for the lowest call priority. Due to the small number of calls, we clustered the calls into two groups: high priority and low priority (See Table~\ref{tab:triage_levels}). We further split the call records into training and test set as shown in Table~\ref{tab:train_test_samples}. 

\subsection{Methods}
%Due to commercial confidentiality we refrain from describing our model and audio features. However, 
We have developed a hierarchical classifier using deep learning neural networks to classify the priority level from the audio representation. Contrary to the traditional approach of analysis of employing audio features such as pitch, articulation rates, etc., we have used deep learning. This is an advanced machine learning technique to learn representation from audio and then use this representation within our deep neural network to classify the call priority. 
%\begin{figure*}[]
%    \centering
%    \rotatebox{90}{\includegraphics[width=\textheight,keepaspectratio]{app_interface.jpg}}
    %\includegraphics[width=1\textwidth]{app_interface.jpg}
%   \caption{Application Interface}
%    \label{fig:app_interface}
%\end{figure*}

\begin{table}[ht]
    %\centering
    \begin{minipage}{0.35\textwidth}
        \centering
\setlength\tabcolsep{0pt}
\begin{tabular}{c >{\bfseries}r @{\hspace{0.7em}}c @{\hspace{0.4em}}c @{\hspace{0.7em}}l}
  \multirow{10}{*}{\rotatebox{90}{\parbox{1.1cm}{\bfseries\centering actual value}}} & 
    & \multicolumn{2}{c}{\bfseries Prediction outcome} & \\
  & & \bfseries p$'$ & \bfseries n$'$ & \bfseries total \\
  & p & \MyBox{35}{} & \MyBox{3}{} & 38 \\[2.4em]
  & n & \MyBox{4}{} & \MyBox{40}{} & 44 \\
  & total & 39 & 43 &
\end{tabular}
        \caption{Confusion Matrix. $p$,$p$$'$ refer to high priority calls and $n$,$n$$'$ refer to low priority calls}
        \label{tab:confusion}
    \end{minipage}%
    \hfill
    \begin{minipage}{0.5\textwidth}
        \centering
        \begin{tabular}{|c|c|}
            \hline
            True Negative Rate/ Specificity  & 91\%  \\ \hline
            True Positive Rate/Recall/ Sensitivity & 92\%  \\ \hline
            False Positive Rate & 9\%  \\ \hline
            False Negative Rate & 8\%  \\ \hline
            Balanced Accuracy  & 92\%  \\ \hline
            Accuracy & 91\%  \\ \hline
            Precision & 90\%  \\ \hline
            F1 score & 91.2\% \\ \hline
        \end{tabular}
        \caption{Results}
        \label{tab:main_results}
    \end{minipage}
\end{table}

\subsection{Results}
Of the 459 calls analysed there were XX males and XX females with a mean age of XX and XX respectively. Table~\ref{tab:demos} below describes these characteristics for each of the triage categories.

\begin{table}[htpb]
\caption{Characteristics of all callers.} 
\label{tab:demos}       
\begin{tabular}{p{0.5cm}|p{2.5cm}|p{1.3cm}|p{1.3cm}|p{1.3cm}|p{1.3cm}|p{1.3cm}|p{1.3cm}|p{1.3cm}|}
    \cline{3-9} 
     \multicolumn{1}{l}{} & \multicolumn{1}{l|}{} & \multicolumn{7}{c|}{\bfseries Triage category \small(n=459)} \\ \cline{3-9} 
        \multicolumn{1}{c}{} & \multicolumn{1}{c|}{} & \multicolumn{4}{c|}{High priority \small(n=212)} & \multicolumn{3}{c|}{ Low priority \small(n=247)} \\
        \cline{3-6} \cline{7-9}   
           \multicolumn{1}{c}{} &  \multicolumn{1}{c|}{}       
            & \centering \small \textbf{A} \newline  Emerg 
            & \centering \small \textbf{B} \newline Very high risk \selectfont 
            &\centering \small \textbf{C} \newline High risk 
            & \centering \small \textbf{D}  \newline Mod risk
            & \centering \small \textbf{E}  \newline Low risk
            & \centering \small \textbf{F} \newline Refer to GP \selectfont
            & \centering \small \textbf{G} \newline Advice \tabularnewline 
            \cline{2-9}
            \parbox[t]{0.5mm}{\multirow{4}{*}{\rotatebox[origin=c]{90}{Male}}} 
            & & & & & & & & \\
            &\small Age Mean(SD) \newline &41(13.2) &37(11.4) &39(13.3) &38(16.5) &35(9.8) &44(12.7) &38(13.1) \\
            \cline{2-9}
           & & & & & & & & \\
           & n(\%) \newline & 15(56\%) &18(58\%) &25(44\%) &35(36\%) &57(37\%) &28(52\%) &20(44\%) \\
            \cline{2-9}
            &&&&&&&&\\
            \parbox[t]{0.5mm}{\multirow{4}{*}{\rotatebox[origin=c]{90}{Female}}} 
            &\small Age Mean(SD) \newline &33(14.3) &32(10.5) &33(11.9) &35(14.5) &35(13.3) &38(16.2) &45(15.7) \\
            \cline{2-9}
            & & & & & & & & \\
            & n(\%) & 12(44\%) &13(42\%) &33(58\%) &61(63\%) &90(58\%) &27(50\%) &25(56\%) \\        
            & & & & & & & & \\
            \cline{2-9}
            &\bfseries Total & 27 & 31 & 58 & 96 & 147 & 55 & 45 \\
            \cline{2-9}

\end{tabular}
\end{table}

The confusion matrix, presented in Table~\ref{tab:confusion}, provides an overview of the model's performance. Out of $38$ high-priority calls, $35$ were accurately classified, while $3$ were misclassified as low priority. Similarly, out of $44$ low-priority calls, $40$ were correctly classified, but $4$ were misclassified as high priority. The model's balanced accuracy, which measures its overall performance, is $92\%$, while its precision is $90\%$.

It's worth noting that there have been no previous studies conducted on mental health helpline data, so we are among the first to achieve these results. Additionally, we aim to minimise the false negative rate, as false positives are more acceptable when it comes to emergency services responses. In this context, false positives refer to low-priority calls being erroneously classified as high-priority, while false negatives indicate high-priority calls being mistakenly classified as low priority. Our false positive rate stands at 9\%, which is low but higher than the false negative rate of 8\%.

\subsection{Application}
\begin{figure*}[ht]
    \centering
    %\rotatebox{90}{\includegraphics[width=\textheight,keepaspectratio]{app_interface.jpg}}
    %\includegraphics[width=1\textwidth]{app_interface.jpg}
    \includegraphics[width=0.83\linewidth]{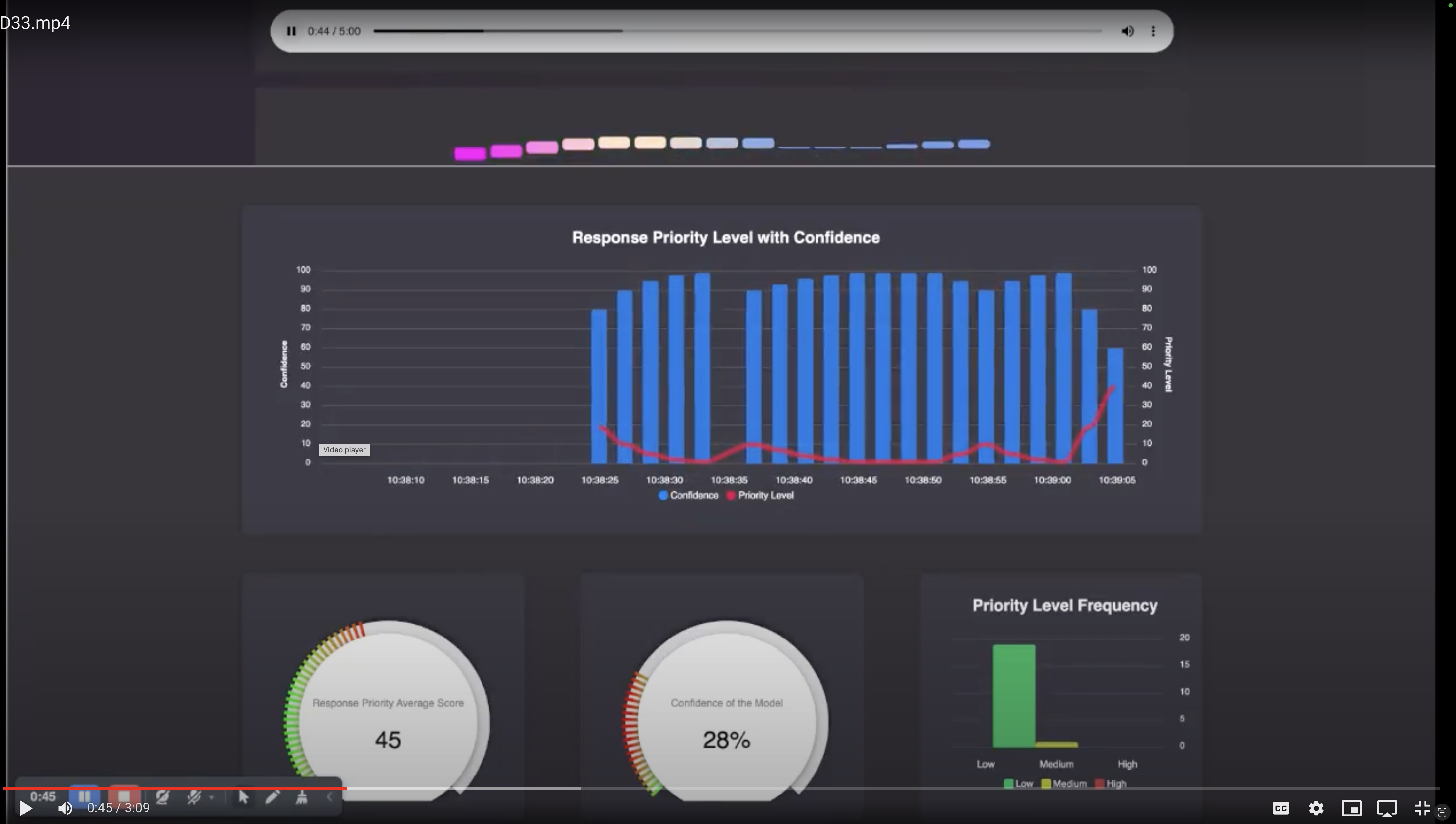}
    \caption{Application Interface}
    \label{fig:app_interface}
\end{figure*}

We have developed an application that implements the algorithms proposed in this paper. An interface of the application is shown in ~Fig.~\ref{fig:app_interface}. This is displayed to the call taker as the call progresses. The very top block shows the audio being played. The next block shows the call priority assessed by the algorithm and the associated confidence in the assessment. The red line shows the priority level, and the corresponding blue bar shows the confidence of the algorithm determining that priority level. The bottom level has two dials and a bar chart. The first dial shows the response priority average score. This keeps updating throughout the call; it might both increase and decrease. The next dial shows the average confidence. As soon as this dial reaches the value above a threshold (say, 90), the priority level shown in the first dial can be used by the call taker as stable information from the algorithm even before the call finishes. However, the algorithm does not make the decision; based on the information provided by the algorithm, the call taker eventually makes the decision. Finally, the third item is the histogram of priority levels, i.e., it shows how many of the audio chunks were high or low priority. 

\section{Discussion}
A significant challenge for clinicians conducting telephone triage is the absence of visual cues. Observation of facial expressions, body language and other behaviours are routinely part of a face-to-face assessment in mental health. Without this crucial element clinicians have an additional challenge in determining the care that should be provided to these callers. Clinicians must rely solely on auditory cues, even if callers do not explicitly mention symptoms like anxiety or depression. It is essential for the clinician to effectively communicate with the caller to accurately assess the priority of potentially life-threatening behaviour. This research has explored the feasibility of implementing a deep learning-based call priority prediction system that utilises voice features rather than the actual content of what is said. The goal of the prototype ML application is to assist clinicians to identify the mental health priority of callers and support their triage decisions. The prototype that has been developed using AI-based technology is intended to streamline tasks such as distress detection, that don’t require a ‘human touch,’ providing complementary support. This allows clinicians to concentrate on delivering more empathic care, thereby ‘humanizing’ clinical practice. Additionally, AI can enhance triage accuracy by supporting the clinical reasoning process, complementing clinical intuition. It is worth noting that since this technology is meant to enhance the clinician's judgement rather than replace it, there remain questions about how to introduce this information into the clinical encounter in a way which is not overly influential on clinician judgements. Therefore the initial interface that was presented in ~Fig.~\ref{fig:app_interface} may not be the final version.

This study demonstrates the feasibility of using machine learning and deep learning techniques to predict call priority in call centres. The proposed models show significant potential in enhancing operational efficiency and patient satisfaction. Most of the process for identifying meaningful patterns in the callers voice data with clinical significance, applicable to day-to-day practice, was achievable through machine learning. By leveraging nonlinear functions and avoiding preconceived assumptions, we have modelled complex relationships from this diverse dataset of over 450 different phone calls to a mental health help line. 

To date there are no AI applications that have been approved or cleared by the therapeutic Goods Administration. However, the concept of such applications have been envisioned for some time, including the potential of expert systems. The delayed adoption of AI into routine mental health practice can possibly be attributed to the sensitive nature of data generated during mental health patient-clinician interactions such as written clinical notes and patient consultations. Additional challenges may also be attributed to the specific and sometimes complex diagnostic criteria defined in the DSM-5 for diagnosis of a psychiatric illness, unlike specific and objective tasks in other fields of healthcare such as oncology to identify a tumour from an image. Additionally, AI technologies require substantial amounts of data, and the field of mental health care has faced limitations in accessing large, high-quality datasets.

The study we conducted here examined whether the emotional features of a caller’s voice could be detected by machine learning algorithms to indicate the caller's level of distress. Results provided an estimate of voice quality on positive and negative demeanour. It did not provide this in real time to the listener and therefore did not influence the outcome of the call. Regarding affective value and ethical considerations, the assessment of AI like our prototype can be seen from two perspectives. On the one hand, this machine learning based AI approach lacks human ethics and reasoning and thereby limits its ability to address affective issues related to mental health. However, alternatively, it is free from the biases that often influence human decisions. As long as our prototype remains confined to its role and does not substitute for mental health professionals, it can serve as an objective and unbiased tool to enhance decision-making.

Addressing ethical considerations in machine learning and deep learning models is crucial. Future work should focus on developing frameworks to ensure that predictive models are fair, transparent, and accountable. It is essential that patients and clinicians have a clear understanding of the reasoning as to what the AI-generated recommendations were based upon. For example, regular audits and adherence to ethical guidelines can help mitigate biases and enhance trust in the system \citep{mittelstadt2019principles}. Clinicians and computer scientists must engage in discussions and debate with consumer representatives and policy makers to help protect against potential abuse of the technology and allow for advancement of this type of technology that could benefit so many in our society.  

% \subsection{Scalability and Performance Optimization}
% As call centres scale, the computational requirements for real-time call priority prediction increase. Research on optimizing model performance and scalability, including the use of distributed computing and efficient model architectures, will be essential to handle large-scale operations \citep{dean2008mapreduce}.

% \section{Results}
% TBA

\subsection{Limitations}
The utilisation of machine learning in assessing the level of distress related to the caller’s mental health has presented unique challenges. The effectiveness of the algorithms used here has relied heavily on the quality of triage categories used for model training. However, due to the inherent heterogeneity of mental illnesses, triage category labels did not provide the specificity needed to achieve high sensitivity and specificity in our AI algorithms. Our alternative approach was to employ ML algorithms for predicting functional consequences in the form of priority level of need rather than focusing solely on risk assessment.

\section{Conclusion}
Future work should focus on refining the models, exploring additional data sources, and addressing implementation challenges. A comprehensive approach involving continuous monitoring and iterative improvements will be essential for successful deployment. Voice computing holds great promise as a clinically assistive tool, but its target clinical applications remain to be fully defined. While our prototype has limitations in providing a true interpretation of a caller’s emotional state, it’s crucial to recognise that our primary goal is not to replace mental health clinicians. Instead, our prototype serves as a supporting source of information for clinicians when they are focused on assessing callers experiencing mental distress. When used judiciously and within appropriate boundaries, it can effectively support mental health services without compromising equitable treatment in understanding and interpreting the caller’s experiences.

%IEEEtran}
\bibliographystyle{plainnat}
\bibliography{main}

\begin{thebibliography}{26}
\providecommand{\natexlab}[1]{#1}
\providecommand{\url}[1]{\texttt{#1}}
\expandafter\ifx\csname urlstyle\endcsname\relax
  \providecommand{\doi}[1]{doi: #1}\else
  \providecommand{\doi}{doi: \begingroup \urlstyle{rm}\Url}\fi

\bibitem[Abi~Kanaan et~al.(2023)Abi~Kanaan, Couchot, Guyeux, Laiymani, Atechian, and Darazi]{kanaan2023methodology}
Marianne Abi~Kanaan, Jean-Fran{\c{c}}ois Couchot, Christophe Guyeux, David Laiymani, Talar Atechian, and Rony Darazi.
\newblock A methodology for emergency calls severity prediction: from pre-processing to bert-based classifiers.
\newblock In \emph{IFIP international conference on artificial intelligence applications and innovations}, pages 329--342. Springer, 2023.

\bibitem[Anthony et~al.(2021)Anthony, Mishra, Stassen, and Son]{anthony2021feasibility}
Tayla Anthony, Amit~Kumar Mishra, Willem Stassen, and Jarryd Son.
\newblock The feasibility of using machine learning to classify calls to south african emergency dispatch centres according to prehospital diagnosis, by utilising caller descriptions of the incident.
\newblock In \emph{Healthcare}, volume~9, page 1107. MDPI, 2021.

\bibitem[Berge et~al.(2023)Berge, Guribye, Fotland, Fonnes, Johansen, and Trattner]{berge2023designing}
Arngeir Berge, Frode Guribye, Siri-Linn~Schmidt Fotland, Gro Fonnes, Ingrid~H Johansen, and Christoph Trattner.
\newblock Designing for control in nurse-ai collaboration during emergency medical calls.
\newblock In \emph{Proceedings of the 2023 ACM Designing Interactive Systems Conference}, pages 1339--1352, 2023.

\bibitem[Botha et~al.(2021)Botha, Morris, Butterworth, and Glozier]{Botha2023}
F.~Botha, R.W. Morris, P.~Butterworth, and N.~Glozier.
\newblock \emph{The kids are not alright: differential trends in mental ill-health in Australia}.
\newblock Melbourne Institute: Applied Economic and Social Research: The University of Melbourne, 2021.

\bibitem[Briggs et~al.(2011)Briggs, Clarke, and Rees]{Briggs2021}
H.~Briggs, S.~Clarke, and N.~Rees.
\newblock Mental health assessment and triage in an ambulance clinical contact centre.
\newblock \emph{Journal of Paramedic Practice}, 13\penalty0 (5):\penalty0 196--203, 2011.

\bibitem[Ceklic et~al.(2022)Ceklic, Ball, Finn, Brown, Brink, Bailey, Whiteside, Brits, and Tohira]{ceklic2022ambulance}
Ellen Ceklic, Stephen Ball, Judith Finn, Elizabeth Brown, Deon Brink, Paul Bailey, Austin Whiteside, Rudolph Brits, and Hideo Tohira.
\newblock Ambulance dispatch prioritisation for traffic crashes using machine learning: a natural language approach.
\newblock \emph{International journal of medical informatics}, 168:\penalty0 104886, 2022.

\bibitem[Defilippo et~al.(2024)Defilippo, Veltri, Li{\'o}, and Guzzi]{defilippo2024leveraging}
Annamaria Defilippo, Pierangelo Veltri, Pietro Li{\'o}, and Pietro~Hiram Guzzi.
\newblock Leveraging graph neural networks for supporting automatic triage of patients.
\newblock \emph{Scientific Reports}, 14\penalty0 (1):\penalty0 12548, 2024.

\bibitem[Downey et~al.(2014)Downey, Zun, and Burke]{Downey2014}
V.~Downey, L.~Zun, and T.~Burke.
\newblock Comparison of emergency nurses association emergency severity triage and australian emergency mental health triage systems for the evaluation of psychiatric patients.
\newblock \emph{Journal of Ambulatory Care Management}, 37\penalty0 (1):\penalty0 11--19, 2014.

\bibitem[Freshworks(2024)]{Freshdesk}
Freshworks.
\newblock Introduction to sentiment analysis: Prioritize tickets based on customer sentiments.
\newblock Available: https://support.freshdesk.com/support/solutions/articles/50000009489-sentiment-analysis-prioritize-tickets-based-on-customer-sentiments, 2024.
\newblock [Accessed Sept 6, 2024].

\bibitem[Genesis(2024)]{GenesisCloud}
Genesis.
\newblock Understand sentiment analysis.
\newblock Available: https://help.mypurecloud.com/articles/understand-sentiment-analysis/, 2024.
\newblock [Accessed Sept 6, 2024].

\bibitem[Holmström et~al.(2011)Holmström, Kaminsky, Spangler, and Winbald]{Holmstrom2020}
I.~Holmström, E.~Kaminsky, D.~Spangler, and U.~Winbald.
\newblock Registered nurses' experiences of using a clinical decision support system for triage of emergency calls: A qualitative interview study.
\newblock \emph{Journal of Advanced Nursing}, 76\penalty0 (11):\penalty0 3104--3112, 2011.

\bibitem[Jahromi et~al.(2024)Jahromi, Ayatollahi, and Ebrazeh]{Jahromi2024}
M.~Jahromi, H.~Ayatollahi, and A.~Ebrazeh.
\newblock Covid-19 hotlines, helplines and call centers: a systematic review of characteristics, challenges and lessons learned.
\newblock \emph{BMC Public Health}, 24:\penalty0 1191, 2024.

\bibitem[Low et~al.(2021)Low, Bentley, and Ghosh]{Low2020}
D.M. Low, K.H. Bentley, and S.S. Ghosh.
\newblock Automated assessment of psychiatric disorders using speech: A systematic review.
\newblock \emph{Laryngoscope Investigative Otolaryngology}, 5:\penalty0 96--116, 2021.

\bibitem[Michel et~al.(2024)Michel, Manns, Boudersa, Jaubert, Dupic, Vivien, Burgun, Campeotto, and Tsopra]{Michel2024}
J.~Michel, A.~Manns, S.~Boudersa, C.~Jaubert, L.~Dupic, B.~Vivien, A.~Burgun, F.~Campeotto, and R.~Tsopra.
\newblock Clinical decision support system in emergency telephone triage: A scoping review of technical design, implementation and evaluation.
\newblock \emph{International Journal of Medical Informatics}, 184:\penalty0 105347, 2024.

\bibitem[Mittelstadt et~al.(2019)Mittelstadt, Russell, and Wachter]{mittelstadt2019principles}
B.~Mittelstadt, C.~Russell, and S.~Wachter.
\newblock Principles alone cannot guarantee ethical ai.
\newblock \emph{Nature Machine Intelligence}, 1\penalty0 (11):\penalty0 501--507, 2019.

\bibitem[Nguyen and Solanki(2021)]{Nguyen2023}
T.P. Nguyen and P.~Solanki.
\newblock Addressing the shortage of psychiatrists in australia: Strategies to improve recruitment among medical students and prevocational doctors.
\newblock \emph{Australian and New Zealand Journal of Psychiatry}, 57:\penalty0 161--163, 2021.

\bibitem[Palaniyappan(2011)]{Palaniyappan2021}
L.~Palaniyappan.
\newblock More than a biomarker: could language be a biosocial marker of psychosis?
\newblock \emph{npj Schizophrenia}, 7\penalty0 (1):\penalty0 42, 2011.

\bibitem[Perez-Siguas et~al.(2023)Perez-Siguas, Matta-Solis, Matta-Solis, Perez-Siguas, Matta-Perez, and Cruzata-Martinez]{perez2023emotion}
Rosa Perez-Siguas, Hernan Matta-Solis, Eduardo Matta-Solis, Luis Perez-Siguas, Hernan Matta-Perez, and Alejandro Cruzata-Martinez.
\newblock Emotion analysis for online patient care using machine learning.
\newblock \emph{Journal of Advanced Research in Applied Sciences and Engineering Technology}, 30\penalty0 (2):\penalty0 314--320, 2023.

\bibitem[Perlini et~al.(2012)Perlini, Marini, Garzitto, Isola, Cerruti, Marinelli, Rambaldelli, Ferro, Tomelleri, Dusi, Bellani, Tansella, Fabbro, and Brambilla]{Perlini2012}
C.~Perlini, A.~Marini, M.~Garzitto, M.~Isola, S.~Cerruti, V.~Marinelli, G.~Rambaldelli, A.~Ferro, L.~Tomelleri, N.~Dusi, M.~Bellani, M.~Tansella, F.~Fabbro, and P.~Brambilla.
\newblock Linguistic production and syntactic comprehension in schizophrenia and bipolar disorder.
\newblock \emph{Acta Psychiatrica Scandinavica}, 126:\penalty0 363--376, 2012.

\bibitem[Sands et~al.(2001)Sands, Elsom, Keppich-Arnold, Henderson, King, Bourke-Finn, and Brunning]{Sands2015}
N.~Sands, S.~Elsom, S.~Keppich-Arnold, K.~Henderson, P.~King, K.~Bourke-Finn, and D.~Brunning.
\newblock Investigating the validity and usability of an interactive computer programme for assessing competence in telephone-based mental health triage.
\newblock \emph{International Journal of Mental Health Nursing}, 25\penalty0 (1):\penalty0 80--86, 2001.

\bibitem[Tanner et~al.(2014)Tanner, Cassidy, and O'Sullivan]{Tanner2014}
R.~Tanner, E.~Cassidy, and I.~O'Sullivan.
\newblock Does using a standardised mental health triage assessment alter nurses assessment of vignettes of people presenting with deliberate self-harm.
\newblock \emph{Advances in Emergency Medicine}, 2014.

\bibitem[Thomas et~al.(2021)Thomas, Schroder, and Tickwood]{Thomas2021}
K.~Thomas, A.~Schroder, and D.~Tickwood.
\newblock A systematic review of current approaches to managing demand and waitlists for mental health services.
\newblock \emph{Mental Health Review Journal}, 25\penalty0 (1):\penalty0 80--86, 2021.

\bibitem[Vigo et~al.(2022)Vigo, Jones, Atun, and Thornicroft]{Vigo2022}
D.~Vigo, L.~Jones, R.~Atun, and G.~Thornicroft.
\newblock The true global disease burden of mental illness: still elusive.
\newblock \emph{Lancet Psychiatry}, 9:\penalty0 98--100, 2022.

\bibitem[Villongco and Khan(2021)]{Villongco2020}
C.~Villongco and F.~Khan.
\newblock “sorry i didn’t hear you.” the ethics of voice computing and ai in high risk mental health populations.
\newblock \emph{AJOB Neuroscience}, 11:\penalty0 105--112, 2021.

\bibitem[Watson et~al.(2022)Watson, Tindall, Patrick, and Moylan]{Watson2022}
T.~Watson, R.~Tindall, A.~Patrick, and S.~Moylan.
\newblock Mental health triage tools: A narrative review.
\newblock \emph{International Journal of Mental Health Nursing}, 32\penalty0 (2):\penalty0 352--364, 2022.

\bibitem[Zendesk(2024)]{Zendesk}
Zendesk.
\newblock The best customer service software for 2024.
\newblock Available: https://www.zendesk.com/service/ticketing-system/customer-service-management-software/, 2024.
\newblock [Accessed Sept 6, 2024].

\end{thebibliography}

\end{document}